\documentclass[12pt]{article}
\usepackage{color}
\usepackage{amsmath}
\usepackage{cite}
\usepackage[pdftex]{graphicx}
\begin{document}    
\vspace*{1cm}

\renewcommand\thefootnote{\fnsymbol{footnote}}
\begin{center} 
  {\Large\bf Scale invariant radiative neutrino mass model}
\vspace*{1cm}

{\Large Daijiro Suematsu}\footnote[1] {professor emeritus, ~e-mail:
suematsu@hep.s.kanazawa-u.ac.jp}
\vspace*{0.5cm}\\

{\it Institute for Theoretical Physics, Kanazawa University, 
Kanazawa 920-1192, Japan}
\end{center}
\vspace*{1.5cm} 

\noindent
{\Large\bf Abstract}\\
We propose a scale invariant radiative neutrino mass model with custodial symmetry 
by introducing two real singlet scalars to the scotogenic model.
Masses of an inert doublet scalar and right-handed neutrinos are induced by a vacuum 
expectation value (VEV) of a singlet scalar caused through the Coleman-Weinberg mechanism.
It violates spontaneously both the custodial symmetry and the scale invariance. 
The weak scale can take a suppressed value compared with the singlet scalar VEV 
because of the custodial symmetry. In this framework we study phenomenological 
consequences for neutrino mass, dark matter and baryon number asymmetry 
by assuming a texture for neutrino Yukawa couplings. 
Since the required dark matter abundance cannot be explained by a neutral component of 
the inert doublet scalar in that case, 
the lightest right-handed neutrino should be dark matter.
Mass of the dark matter is predicted to be less than $O(1)$ MeV and baryon number asymmetry
could be explained through resonant leptogenesis.   

\newpage
\setcounter{footnote}{0}
\renewcommand\thefootnote{\alph{footnote}}

\section{Introduction}
The standard model (SM) is a successful theoretical framework which can explain experimental 
results known by now although some unsatisfactory points remain in it.
The fatal ones among them are that it cannot explain both existence of neutrino masses and 
dark matter (DM) and also the asymmetry of matter and antimatter in the Universe. 
These are confirmed through various experiments and observations \cite{noscil,dm,basym}.
It suggests that the SM should be extended so as to explain them at least.
Since the SM has a robust structure, its extension seems to be desirable to give a simultaneous 
explanation for them without causing any contradictions.

As such an example, we have the scotogenic model \cite{ma} which is a simple extension of 
the SM with an inert doublet scalar $\eta$ and three right-handed neutrinos $N_j$. 
If these new contents are assumed to have a $Z_2$ odd charge and the SM contents have its even 
charge, these three problems can be simultaneously solved under the assumption that
the $Z_2$ symmetry is exact \cite{scot}.
Unfortunately, the model cannot give any explanation for mass scales contained in it.
The model is characterized by three mass scales, that is, the ordinary weak scale whose origin is 
a vacuum expectation value (VEV) of the Higgs scalar, the mass of the inert doublet scalar and 
the one of the right-handed neutrinos. 
A typical feature of the model is that the latter two scales could be taken at a TeV region. 
However, these are fixed by hand and the well-known hierarchy problem is 
left as an untouchable problem in the model. 
     
Recently, an interesting solution for the hierarchy problem of the SM has been proposed 
as an extended model based on both the classical scale invariance and the custodial 
symmetry \cite{cust1,cust2}.
In the model given in \cite{cust2}, the scalar sector is extended by two real singlet scalars, and 
the Higgs doublet scalar is assumed to compose a {\bf 5}-plet of custodial symmetry $SO(5)$
with one of these singlet scalars. 
The Higgs scalar behaves as Nambu-Goldstone bosons due to the spontaneous 
breaking of this custodial symmetry at an intermediate scale.
The breaking is caused by a VEV of the singlet scalar due to the Colman-Weinberg mechanism \cite{cw}. 
Since negative squared Higgs mass can be induced radiatively through the explicit breaking of 
the custodial symmetry, the weak scale could be much smaller than the intermediate scale
thanks to the custodial symmetry.
   
In this paper, we consider to apply this idea to the scotogenic model by introducing two $Z_2$ 
even singlet scalars. This extended model may be considered as a UV model 
which can give a solution to the problem in  the scotogenic model.  
The above mentioned two mass scales, that is, the inert doublet scalar mass 
and the right-handed neutrino mass, are considered to be generated through the couplings 
with the singlet scalar which causes the spontaneous breaking of the custodial symmetry. 
Since these couplings violate the custodial symmetry explicitly, they have to be constrained 
so that the weak scale is generated as an appropriately suppressed one. 
Quartic couplings of the inert doublet scalar in the scalar potential are also constrained by the same reason
although they play a crucial role to reduce the DM abundance if DM is a neutral 
component of the inert doublet scalar.
These suggest that phenomenological consequences could be largely changed from the
usual ones of the scotogenic model.
As such examples, we discuss DM and leptogenesis by assuming 
a texture of neutrino Yukawa couplings. We find that DM should be a lightest right-handed neutrino.  
Mass degeneracy among the inert doublet scalar and the right-handed neutrinos which are required for 
the suitable generation of the weak scale makes the resonant leptogenesis work well. 

The remaining parts are organized as follows. In section 2, we define the model studied in this paper. 
We address how the weak scale can be induced appropriately and what constraints should be imposed
on the custodial symmetry violating couplings.  
In section 3, we discuss consequences caused by such constraints for
the neutrino mass generation, the DM abundance and the baryon number 
asymmetry through leptogenesis. We summarize the paper in section 4. 

\section{Scale invariant scotogenic model}
A model considered in this paper is an extension of the scotogenic model with two real singlet 
scalars $\phi$ and $S$. 
It is assumed to have $Z_2$ invariance for which the inert doublet scalar $\eta$ and 
the right-handed neutrinos $N_j$ are assumed to have its odd charge 
but two new singlet scalars have its even charge as the SM contents.
We also impose on the model both the classical scale invariance and the custodial symmetry 
$SO(5)$. The Higgs doublet scalar $H$ and $\phi$ are assumed to 
transform as a ${\bf 5}$-plet of this $SO(5)$.

The model is defined by renormalizable scalar potential $V$ and Yukawa couplings of $N_j$ 
in the leptonic sector as
\begin{eqnarray}
V&=&\lambda_{c1}\left[(H^\dagger H)+\frac{1}{2}\phi^2\right]^2
+\frac{1}{2}\lambda_{c2}\left[(H^\dagger H)+\frac{1}{2}\phi^2\right]S^2 +\frac{1}{4!}\lambda_SS^4
+\lambda_\eta(\eta^\dagger\eta)^2 \nonumber \\
&+&\frac{1}{2}\lambda_{\eta \phi}\phi^2(\eta^\dagger\eta)+
\frac{\tilde\lambda_5}{2}\Big[(H^\dagger\eta)^2+(\eta^\dagger H)^2\Big], \nonumber \\
-{\cal L}_Y&=&\sum_{j=1}^3\left(\sum_{\alpha=e,\mu,\tau}h_{\alpha j}\bar\ell_\alpha\eta N_j 
+\frac{1}{2}y_{N_j}\phi\bar N_jN_j^c\right)+{\rm h.c.},
\label{custlag}
\end{eqnarray}
where $\ell_\alpha$ is a left-handed doublet lepton.
 Quartic scalar couplings $S^3\phi$ and $S\phi^3$ are not included in the scalar potential
since they violate the custodial symmetry. Both quartic couplings 
$\tilde\lambda_3(H^\dagger H)(\eta^\dagger\eta)$ and 
$\tilde\lambda_4(H^\dagger\eta)(\eta^\dagger H)$ in the ordinary scotogenic model 
are not included since they also violate it. On the other hand,   
the custodial symmetry is explicitly violated by gauge interactions and Yukawa couplings in the SM.  
Although quartic couplings in the second line of $V$ and Yukawa couplings 
$y_{N_j}\phi \bar N_jN_j^c$ of the right-handed neutrinos violate it, they are included 
as important terms.
These couplings not only determine the mass of $\eta$ and $N_j$ 
through the VEV of $\phi$ but also play crucial roles for the generation of the weak 
scale as shown later. 
If we rewrite the potential $V$ as
\begin{eqnarray}
V_0&=&\lambda_H(H^\dagger H)^2+\lambda_{H\phi}\phi^2(H^\dagger H)+\frac{1}{4}\lambda_\phi\phi^4
+\frac{1}{2}\lambda_{HS}S^2(H^\dagger H) 
+\frac{1}{4}\lambda_{\phi S}\phi^2S^2 \nonumber \\
&+&\frac{1}{4!}\lambda_SS^4+\lambda_\eta(\eta^\dagger\eta)^2
+\frac{1}{2}\lambda_{\eta \phi}\phi^2(\eta^\dagger\eta)
+\frac{\tilde\lambda_5}{2}\Big[(H^\dagger\eta)^2+(\eta^\dagger H)^2\Big],
\label{spot}
\end{eqnarray}
the custodial symmetry imposes the conditions 
\begin{equation}
\lambda_H=\lambda_{H\phi}=\lambda_\phi=\lambda_{c1}, \qquad 
\lambda_{HS}=\lambda_{\phi S}=\lambda_{c2} 
\label{init}
\end{equation}
at a cut-off scale which we take as the Planck scale in the present study.
Conditions $\lambda_H,\lambda_\phi, \lambda_S,\lambda_\eta>0$ should be satisfied for the
stability of the potential at least.

If the spontaneous symmetry breaking of the custodial symmetry is caused by a nonzero VEV 
of $\phi$ due to quantum effects at some intermediate scale, 
$\lambda_\phi$ should be small enough at that scale. 
Such a situation is expected to happen in a case where $V_0$ 
has a flat direction determined mainly by $\phi$ \cite{flat}. 
Here, we consider background fields of the scalars which are represented 
as $S_b, \eta_b, \phi_b$ and $H_b$.  Components of the last one are 
expressed as $\frac{1}{\sqrt 2}H_{bi}~(i=1-4)$.
In the case $S_b=\eta_b=0$,  $V_0$ has a flat direction determined by $\phi_b$ and $H_b$ under a condition
\begin{equation}
2\lambda_H(H_b^\dagger H_b)+\lambda_{H\phi}\phi_b^2=0, \qquad 
\lambda_H=\frac{\lambda_{H\phi}^2}{\lambda_\phi}.
\label{flat}
\end{equation}
This condition requires $\lambda_{H\phi}<0$ and the first one may be represented as
\begin{equation}
H_{b1}=H_{b2}=H_{b3}=0, \quad H_{b4}=\sqrt\frac{-\lambda_{H\phi}}{\lambda_H-\lambda_{H\phi}}v, 
\quad \phi_b=\sqrt\frac{\lambda_H}{\lambda_H-\lambda_{H\phi}}v,
\label{fvev}
\end{equation}
where $v$ is defined as $v=\sqrt{2H_b^\dagger H_b+\phi_b^2}$.
If $\lambda_H\gg |\lambda_{H\phi}|$ is satisfied, $H_{b4}$ can be much smaller than $\phi_b$. 

We focus our study on this flat direction defined by nonzero background fields $\phi_b$ and $H_{b4}$. 
The fields on this background can get mass depending on them.
If we define $x$ as $x\equiv H_{b4}/\phi_b$, mass eigenvalues of six scalar contents 
are found to be
\begin{equation}
(\lambda_{H\phi}+\lambda_Hx^2)\phi_b^2, \qquad 
\frac{1}{2}(\lambda_{\phi S}+\lambda_{HS}x^2)\phi_b^2
\end{equation}  
for $H_{1,2,3}$ and $S$, and also the ones of the mass matrix 
\begin{equation}
\left(\begin{array}{cc}3\lambda_\phi+\lambda_{H\phi}x^2 & 2\lambda_{H\phi}x \\
2\lambda_{H\phi}x & \lambda_{H\phi}+3\lambda_Hx^2
 \end{array}\right)\phi_b^2
\label{dhmatrix}
\end{equation}
for $(\phi, H_4)$.
Mass of the inert doublet $\eta$ and the right-handed neutrino $N_j$ is respectively given as
\begin{equation} 
m_\eta^2=\frac{1}{2}\lambda_{\eta\phi}\phi_b^2, \qquad M_j=y_{N_j}\phi_b,
\label{etan-mass}
\end{equation}
where a contribution from the $\tilde\lambda_5$ term is not explicitly written in the formula of 
 $m_\eta^2$. 
Using these field dependent mass eigenvalues, the one-loop effective potential can be expressed as
\begin{equation}
V_{\rm eff}=V_0+V_1, \qquad 
V_1=\sum_i\frac{n_i(-1)^{2s_i}}{64\pi^2}m_i^4\left(\ln\frac{m_i^2}{\mu^2}-\frac{3}{2}\right),
\end{equation}
where $n_i$ and $s_i$ are a degree of freedom and a spin of the field $i$, respectively.

We are interested in a vacuum of the model which satisfies $x\ll 1$.
Such a vacuum can be determined as a minimum of the effective potential $V_{\rm eff}$ 
which can be consider as a function of $H_b$ and $\phi_b(H_b)$ \cite{cust1,cust2}. 
It can be found by solving the potential minimum condition
\begin{equation}
\frac{dV_{\rm eff}}{dH_b}=\frac{\partial V_{\rm eff}}{\partial H_b}+\frac{d\phi_b}{dH_b}
\frac{\partial V_{\rm eff}}{\partial \phi_b}=0.
\end{equation}
By solving this equation for $\phi_b$ at $H_b=0$, we find a solution $\phi_0$ under an assumption 
$\lambda_{1c}\ll \lambda_{2c}$ as
\begin{equation}
\ln\frac{\phi_0^2}{\mu^2}
=\frac{-64\pi^2\lambda_\phi
-\lambda_{\phi S}^2\left(\ln\frac{\lambda_{\phi S}}{2}-1\right)
+{\cal A}}
{\lambda_{\phi S}^2+16\lambda_{H\phi}^2+36\lambda_\phi^2+4\lambda_{\eta\phi}^2
-8\sum_j y_{N_j}^4} \simeq-64\pi^2\frac{\lambda_\phi}{\lambda_{\phi S}^2}
-\ln\frac{\lambda_{\phi S}}{2}+1,
\label{phivev}
\end{equation} 
where ${\cal A}$ is represented as\footnote{Since this expression becomes complex for 
$\lambda_{H\phi}<0$, one might suspect the validity of eq.~(\ref{phivev}) in that case.
However, if $\mu$ is fixed as a value where $\lambda_{H\phi}$ changes its sign from positive to 
negative, it can give a consistent and stable result. In fact, since the corresponding term 
is smaller than others by several orders of magnitude in ${\cal A}$ around such a $\mu$, 
this can be a good prescription for the problem. }  
\begin{equation}
{\cal A}=-16\lambda_{H\phi}^2\left(\ln\lambda_{H\phi}-1\right)
-36\lambda_\phi^2\left(\ln 3\lambda_\phi-1\right)-4\lambda_{\eta\phi}^2
\left(\ln \frac{\lambda_{\eta\phi}}{2}-1\right)
+8\sum_j y_{N_j}^4\left(\ln y_{N_j}^2-1\right) \nonumber.
\end{equation}
If $\eta$ and $N_{2,3}$ have nearly degenerate mass, eq.~(\ref{etan-mass}) suggests 
$\lambda_{\eta\phi}\sim 2y_{N_{2,3}}^2$ and then the last two terms in both ${\cal A}$ 
and a denominator of eq.~(\ref{phivev}) cancel each other in the case 
$y_{N_1}\ll y_{N_{2,3}}$.\footnote{We discuss the assumption $M_1\ll M_{2,3}$ in the part 
for the mass generation of active neutrinos later.}
In that case, their contribution could be suppressed to the same order or less compared 
with the one from $\lambda_{\phi S}$ when $\lambda_\phi, |\lambda_{H\phi}|\ll\lambda_{\phi S}$ 
is satisfied. 
We can expect that the VEV $v_\phi$ of $\phi$ described by $\phi_0$ could take a value in TeV
regions if couplings $\lambda_\phi$ and $\lambda_{\phi S}$ are fixed appropriately. 

In order to determine a VEV of $H$, we consider a perturbative expansion 
of the effective potential $V_{\rm eff}$ with respect to $x$ around the minimum 
$\phi_0$ given in eq.(\ref{phivev}). 
For this purpose, we consider to expand it by $\epsilon$ which is formally introduced 
by replacing of $x$ and $\lambda_{H\phi}$ with $\epsilon x$ 
and $\epsilon^2\lambda_{H\phi}$, respectively \cite{cust1,cust2}. 
This expansion guarantees the condition (\ref{flat}) since this replacement keeps it unchanged.
We pick up higher order terms of $x$ by applying this expansion method up to $\epsilon^4$ terms.  
As a result of this procedure, we can derive the effective scalar potential 
${\cal V}$ at $\mu$ which includes the scalar potential of the ordinary 
scotogenic model as its part. 
Such scalar potential ${\cal V}$ is found to be written as
\begin{eqnarray}
{\cal V}&=&\left(\lambda_{H\phi}-\frac{\lambda_{HS}}{\lambda_{\phi S}}\lambda_\phi\right)v_\phi^2
(H^\dagger H)+\left(\lambda_H  
-\frac{\lambda_{H S}^2}{\lambda_{\phi S}^2}\lambda_\phi+
\frac{\lambda_{HS}^2}{64\pi^2}\right)(H^\dagger H)^2 \nonumber \\ 
&+&\lambda_\eta(\eta^\dagger\eta)^2 +\frac{\tilde\lambda_5}{2}
\Big[(H^\dagger\eta)^2+(\eta^\dagger H)^2\Big]
+\frac{\lambda_{\eta\phi}}{2}v_\phi^2(\eta^\dagger\eta)
+\sum_k \frac{n_k (-1)^{2s_k} }{64\pi^2}m_k^4\left(\ln\frac{m_k^2}{\mu_0^2}-C_k\right)\nonumber \\
&-&\frac{\lambda_{\phi S}^2}{512\pi^2}v_\phi^4+
\frac{3\lambda_{\phi S}^2}{512\pi^2} \phi^4+\frac{1}{2}\tilde\beta_{\lambda_\phi}v_\phi^2\phi^2
+ 2\left( \lambda_{H\phi} -\frac{\lambda_{HS}} {\lambda_{\phi S}} \lambda_\phi
+\frac{1}{64\pi^2}\lambda_{\phi S}\lambda_{HS} \right) v_\phi\phi (H^\dagger H) \nonumber \\
&+&\left(\lambda_{H\phi}+\frac{7}{64\pi^2}\lambda_{HS}\lambda_{\phi S}
\right)\phi^2(H^\dagger H)
+\frac{1}{4!}\lambda_SS^4+ \frac{1}{2}\lambda_{\phi S}(v_\phi+\phi)^2S^2
+\frac{1}{2}\lambda_{HS}(H^\dagger H)S^2, \nonumber\\
\label{effpot}
\end{eqnarray}
where $\phi$ is redefined as $v_\phi+\phi$ and couplings between $\phi$ and $\eta$
are abbreviated. $\tilde\beta_{\lambda_\phi}$ equals to the 
$\beta$-function of the quartic coupling $\lambda_\phi$ given in the Appendix 
except for the contribution from the anomalous dimension of $\phi$.
The summation for $k$ should be done for the SM contents and $C_k=\frac{3}{2} (\frac{5}{6})$ 
for scalars and fermions (vector bosons).
Since the first two lines of ${\cal V}$ are considered to correspond to the scalar potential 
of the ordinary scotogenic model given as\footnote{It should be noted that the present 
low energy effective model is not the ordinary scotogenic model but it is extended by the 
additional couplings of $\phi$ and $S$ from the ordinary scotogenic model as found in   
${\cal V}$. However, they do not affect the important formulas (\ref{etamass1}) and 
(\ref{nmass}) for the inert doublet scalar mass and the neutrino mass in the scotogenic model. }
\begin{eqnarray}
V_{\rm scot}&=&m_H^2H^\dagger H +\tilde\lambda_1(H^\dagger H)^2+\tilde\lambda_2(\eta^\dagger\eta)^2
+\tilde\lambda_3(H^\dagger H)(\eta^\dagger\eta)+\tilde\lambda_4(H^\dagger\eta)(\eta^\dagger H)
\nonumber \\
&+&\frac{\tilde\lambda_5}{2}\Big[(H^\dagger \eta)^2+(\eta^\dagger H)^2\Big]
 +m_\eta^2\eta^\dagger\eta +\sum_k \frac{n_k (-1)^{2s_k} }{64\pi^2}
m_k^4\left(\ln\frac{m_k^2}{\mu_0^2}-C_k\right),
\label{scot}
\end{eqnarray}
we find the parameters in eq.~(\ref{scot}) can be expressed by using the ones 
in eq.~(\ref{spot}) at $\mu$ as follows,
\begin{eqnarray}
&&m_H^2\simeq\left(\lambda_{H\phi}-\frac{\lambda_{H S}}{\lambda_{\phi S}}\lambda_\phi\right)v_\phi^2, 
\quad \tilde\lambda_1\simeq \lambda_H  
-\frac{\lambda_{H S}^2}{\lambda_{\phi S}^2}\lambda_\phi+\frac{\lambda_{HS}^2}{64\pi^2}, \quad
m_\eta^2=\frac{1}{2}\lambda_{\eta\phi}v_\phi^2,   \nonumber \\ 
&&\tilde\lambda_2=\lambda_\eta, \qquad \tilde\lambda_3\simeq 0, \qquad \tilde\lambda_4\simeq 0.
\label{scotpara}
\end{eqnarray}
Nonzero $\tilde\lambda_3$ and $\tilde\lambda_4$ could be induced radiatively 
at this scale even if they are assumed to be zero at the cut-off scale.
However, their values are found to be small enough and then we can suppose them to be zero safely.
Mass of the right-handed neutrino $N_j$ is given in eq.~(\ref{etan-mass}).

If both $\lambda_\phi>0$ and $\lambda_{H\phi}<\lambda_\phi$ are satisfied at some critical scale,
the first condition in (\ref{scotpara}) shows that $m_H^2<0$ could be realized to induce the spontaneous 
electroweak symmetry breaking.  
The difference $|\lambda_{H\phi}-\frac{\lambda_{HS}}{\lambda_{\phi S}}\lambda_\phi|$ could be substantially 
suppressed to result in $|m_H^2|\ll v_\phi^2$ because of the custodial symmetry which guarantees 
$\lambda_{H\phi}\simeq \lambda_\phi$ and  $\lambda_{HS}\simeq \lambda_{\phi S}$.
The Higgs scalar in this model is not a pure SM Higgs scalar.  As shown in eqs.~(\ref{dhmatrix}) and
(\ref{effpot}), 
it has a small mixing with $\phi$.  
If we write the mass eigenstates as $h$ for a Higgs-like scalar and $h_\phi$, their mass eigenvalues are 
found through $\partial^2 {\cal V}/\partial f_i\partial f_j$. They are written as
\begin{equation}
m_h^2\simeq -2\left(\lambda_{H\phi}-\frac{\lambda_{HS}}{\lambda_{\phi S}}\lambda_\phi \right)v_\phi^2, 
\qquad
m_{h_\phi}^2\simeq \frac{\lambda_{\phi S}^2}{32\pi^2}v_\phi^2,
\label{etam}
\end{equation}
and their mixing angle $\theta$ is expressed as 
\begin{equation}
\tan\theta\simeq 
\frac{2\left(\lambda_{H\phi}-\frac{\lambda_{HS}}{\lambda_{\phi S}}\lambda_\phi
+\frac{\lambda_{HS}\lambda_{\phi S}}{64\pi^2}\right) v_\phi v_H}
{m_{h_\phi}^2-m_h^2}.
\end{equation}
Since $m_{h_\phi}^2$ is determined by the $\beta$-function of the quartic 
coupling $\lambda_\phi$, $h_\phi$ can be identified with a dilaton, a psudo Nambu-Goldstone 
boson caused by the spontaneous violation of the classical scale invariance. 
The Higgs VEV $\langle H\rangle(\equiv v_H/\sqrt 2)$ can be estimated through a relation 
$v_H=m_h/\sqrt{2\tilde\lambda_1}$ by using $\tilde\lambda_1$ and $m_h$.

Now, we examine this scenario numerically by fixing the parameters in $V_0$ 
at the Planck scale. In this paper, we pick up benchmark parameters as an example and 
present physical results for them to describe features of the model.  
As such benchmark parameters, we take them at the Planck scale as follows, 
\begin{eqnarray}
&&\lambda_{c1}=4.48\times 10^{-3}, \quad \lambda_{c2}=2.27\times 10^{-1}, \quad 
\lambda_S=0.2 , \quad \lambda_\eta=0.19,   \nonumber \\ 
&&  \tilde\lambda_5=-3\times 10^{-3},  \quad 2\lambda_{\eta\phi}=(y_{N_2}+y_{N_3})^2, 
\quad y_{N_2}= y_{N_3}=2.03\times 10^{-1},    
\label{bench}
\end{eqnarray}
where $y_{N_1}\ll y_{N_2}$ is assumed.  Couplings which violate the custodial symmetry are listed 
in the second line.\footnote{A simple relation is assumed between $\lambda_{\eta\phi}$ 
and $y_{N_{2,3}}$ so that $m_\eta$ and $M_{2,3}$ take a degenerate value at a low energy scale.
$\lambda_S$ and $\lambda_\eta$ should be fixed so as to guarantee their positivity 
at the intermediate scale.}
Our observations addressed above can be confirmed by solving the renormalization group equations 
(RGEs) of the relevant couplings for these initial values. RGEs used in this analysis are 
presented in Appendix.

\begin{figure}[t]
\vspace*{-1cm}
\begin{center}
\includegraphics[width=9cm]{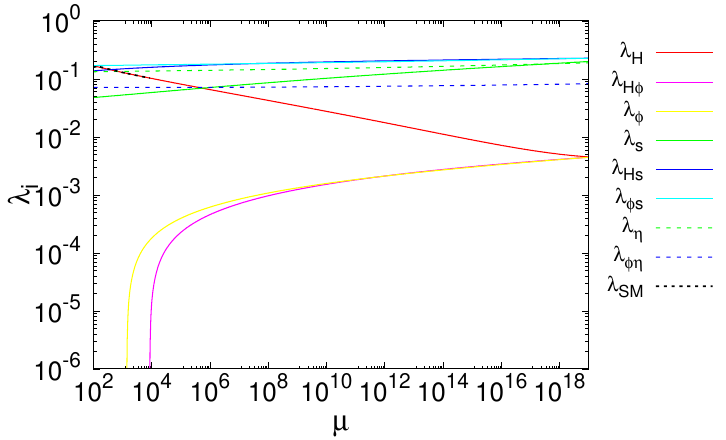}
\end{center}
\vspace*{-5mm}
\footnotesize{{\bf Fig.~1}~~Solutions of the RGEs for the quartic scalar couplings in the model.
The parameters given in eq.~(\ref{bench}) are used as the initial values at the Planck scale.  
$y_{N_{2,3}}$ and $\lambda_{\eta\phi}$ fixed at the Planck scale in eq.~(\ref{bench}) induce 
the nearly degenerate mass for $\eta$ and $N_{2,3}$ at $\mu\simeq M_2$. 
The result is confirmed to be stable for $y_{N_1}$ if its initial value is taken less than $10^{-4}$.  }
\end{figure}

In Fig.~1, we plot solutions of the RGEs  of the quartic scalar couplings in eq.~(\ref{spot}).
In this plot, we use eq.~(\ref{h23}) for the neutrino Yukawa couplings $h_{\alpha j}$ which are 
fixed through the neutrino oscillation data as discussed in the next part. Their values at the 
Planck scale are determined from them.
The figure shows that $\lambda_\phi$ takes a small positive value at a critical scale 
$\mu\simeq 8800$ GeV where $\lambda_{H\phi}$ changes its sign. 
On the other hand, $\lambda_H$ takes a much larger value than $|\lambda_{H\phi}|$ 
thanks to the top Yukawa coupling, which is favored as addressed below eq.~(\ref{fvev}). 
Since $\lambda_{H\phi}-\frac{\lambda_{H S}}{\lambda_{\phi S}}\lambda_\phi<0$ is realized and its absolute value is sufficiently suppressed, we find from eq.(\ref{scotpara}) that 
a desirable $m_H^2$ could be generated to cause 
a VEV of the Higgs scalar such as  $v_H \ll v_\phi$.
Using the solutions in Fig.~1,  we find $v_\phi\simeq 7880$ GeV from eq.~(\ref{phivev})
and rather good numerical values are obtained for the Higgs mass, the Higgs VEV, 
the top mass, and the dilaton mass such that 
\begin{equation}
 \qquad m_h\simeq 137~{\rm GeV}, \qquad 
v_H \simeq 242~{\rm GeV}, \qquad m_t\simeq 175~{\rm GeV}, \qquad m_{h_\phi}\simeq 77~{\rm GeV}.
\end{equation}
The mixing between the Higgs-like boson and the dilaton is estimated as $\tan\theta\simeq 0.04$.
The masses of $\eta$,  $N_{2,3}$ and $S$ are found to be
\begin{equation}
m_\eta\simeq 1480~{\rm GeV}, \qquad M_{2,3}\simeq 1500~{\rm GeV}, \qquad m_S\simeq 2320~{\rm GeV}.
\label{mscalar}
\end{equation} 
Small mass difference of  $N_2$ and $N_3$ is induced from the neutrino Yukawa couplings $h_{\alpha j}$
through the RGE evolution of $y_{N_j}$. It is estimated as $(M_2-M_3)/M_2=8.8\times 10^{-7}$ 
at their mass scale.

\begin{figure}[t]
\vspace*{-1cm}
\begin{center}
\includegraphics[width=8cm]{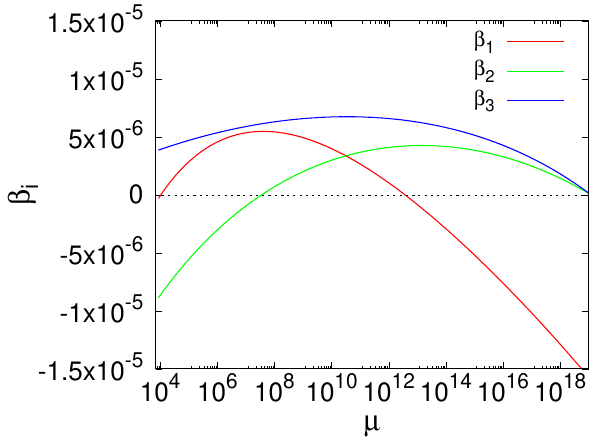}
\end{center}
\vspace*{-5mm}
\footnotesize{{\bf Fig.~2}~~~Evolution of $\beta_1,~\beta_2$ and $\beta_3$ in eq.~(\ref{beta}), 
whose definition is given in the text. 
The initial condition $\lambda_{c1}\ll\lambda_{c2}$ allows $\beta_2$ to play an important role
in spite of the custodial symmetry. The behavior of the $\beta_1$ is considered to be caused by the top Yukawa coupling. } 
\end{figure}

It may be instructive to examine how $|m_H^2| \ll v_\phi^2$ can be realized more closely.
The difference $\lambda_{H\phi}-\frac{\lambda_{HS}}{\lambda_{\phi S}}\lambda_\phi$ at a critical scale is determined 
by the difference of their $\beta$-functions. If we take account of the constraint imposed by 
the custodial symmetry, it could be estimated as
\begin{eqnarray}
\beta_{\lambda_{H\phi}}-\beta_{\lambda_\phi}&\simeq& \frac{1}{16\pi^2}\left[
\lambda_{H\phi}\left(6y_t^2+2\sum_jy_{N_j}^2+12\lambda_H-\frac{3}{2}g_Y^2-\frac{9}{2}g_2^2\right)
-4\lambda_\phi\sum_jy_{N_j}^2\right. \nonumber \\
&+&\frac{1}{2}\lambda_{\phi S}(\lambda_{HS}-\lambda_{\phi S})
+\left.4\sum_j y_{N_j}^4-2\lambda_{\eta\phi}^2\right]\equiv \beta_1+\beta_2+\beta_3,
\label{beta}
\end{eqnarray}
where $\beta_1$, $\beta_2$ and $\beta_3$ correspond to the first line, 
the first term in the second line, and the last two terms in the second line, respectively.
We should note that $\beta_2$ cannot be neglected even under the constraint of 
the custodial symmetry since it could give the same order contribution as the one from $\beta_1$ 
under the initial condition $\lambda_{c2}\gg\lambda_{c1}$ assumed in eq.~(\ref{bench}). 
On the other hand, since $\beta_3$ comes from the terms which break the custodial symmetry,
it is not controlled by the symmetry so that there is no reason why its contribution can be 
sufficiently suppressed.
Here, we should note that two terms in $\beta_3$ are relevant to 
the masses of $N_j$ and $\eta$ which are generated through the VEV of $\phi$, respectively. 
If the mass of $N_{2,3}$ is nearly degenerate with the one of $\eta$ and then 
$2y_{N_{2,3}}^2\simeq\lambda_{\eta\phi}$ is satisfied at low energy regions,\footnote{
This condition is found to play a crucial role in the study of dark matter and leptgenesis as seen later. }  
its contribution could be constrained.

Assuming this mass degeneracy and using the parameters in eq.~(\ref{bench}), 
we plot the evolution of $\beta_1,~\beta_2$ and $\beta_3$ in Fig.~2.
The figure shows that the magnitude of $\beta_3$ is comparable to the one of $\beta_1$ and $\beta_2$ throughout the relevant energy scale.
Both $\beta_2$ and $\beta_3$ are found to play a crucial role to cancel the negative 
contribution of $\beta_1$ at high energy regions. It makes 
$\lambda_{H\phi}-\frac{\lambda_{HS}}{\lambda_{\phi S}}\lambda_\phi$ possible to take 
a negative and appropriate value at low energy regions.

This analysis shows that the present model can induce three mass scales of the 
scotogenic model in a desirable way as long as suitable terms for explicit breaking of the custodial 
symmetry are introduced.
In the next section, we study phenomenological consequences caused by the constraints on 
the parameters of the scotogenic model given in eq.(\ref{scotpara}).
We focus our attention on their influence to neutrino mass, dark matter and leptogenesis for 
the baryon number asymmetry. 
Numerical study in the following part is based on the benchmark parameters presented 
in  eq.(\ref{bench}) and the results of the RGEs analysis in this section.  

\section{Phenomenological consequences}
\subsection{Neutrino mass}
The inert doublet scalar $\eta$ is a key ingredient of the neutrino mass generation 
in the scotogenic model.
Mass of its components is expressed as
\begin{equation}
M_{\eta^\pm}^2=m_\eta^2+\frac{1}{2}\tilde\lambda_3v_H^2, \qquad
M_{^{\eta_R}_{\eta_I}}^2= m_\eta^2+\frac{1}{2}(\tilde\lambda_3+\tilde\lambda_4\pm\tilde\lambda_5)v_H^2,
\label{etamass1}
\end{equation}
where $\eta^\pm$ are  charged components and $\eta_{R(I)}$ is a real (imaginary) part 
of the neutral one, respectively.  
In the present extended model, $\tilde\lambda_3$ and $\tilde\lambda_4$ are assumed to be zero 
at the Planck scale. Although they take nonzero values radiatively through RGE effects of 
the small custodial symmetry breaking such as the $\tilde\lambda_5$ coupling assumed 
in eq.~(\ref{bench}), they are found to be negligibly small.
Taking account of this point and the expression of $m_\eta^2$ given in eq.~(\ref{scotpara}), 
the mass of each component of $\eta$ is found to be represented as  
\begin{equation}
M_{\eta^\pm}^2\simeq \frac{1}{2}\lambda_{\eta\phi}v_\phi^2, \qquad 
M_{^{\eta_R}_{\eta_I}}^2\simeq \frac{1}{2}\lambda_{\eta\phi}v_\phi^2\pm\frac{1}{2}\tilde\lambda_5v_H^2.
\label{etamass2}
\end{equation}

Neutrino mass in the scotogenic model is generated through one-loop diagrams 
which have a neutral component of $\eta$ and a right-handed neutrino $N_j$ 
in the internal lines. A mass matrix caused by them can be expressed as 
\begin{equation}
({\cal M}_\nu)_{\alpha\beta}=\sum_{j=1}^3 h_{\alpha j} h_{\beta j}\Lambda_j, \quad 
\Lambda_j=\frac{\tilde\lambda_5v_H^2}{32 \pi^2M_j}
  \frac{M_j^2}{M_\eta^2-M_j^2} \left(1+\frac{M_j^2}{M_\eta^2-M_j^2}
    \ln\frac{M_j^2}{M_\eta^2}\right) ,
\label{nmass}
\end{equation}
where $M_\eta^2=\frac{1}{2}(M_{\eta_R}^2+M_{\eta_I}^2)$.
As found from this formula, $\tilde\lambda_5\not=0$ is a necessary condition of 
the neutrino mass generation in this scheme.
Since a small $\tilde\lambda_5$ term is included in eq.~(\ref{custlag}) as one of the custodial 
symmetry violating terms,\footnote{It should be noted that 
$\tilde\lambda_5$ is irrelevant to both $\beta_{\lambda_{H\phi}}$ and $\beta_{\lambda_\phi}$
in eq.~(\ref{beta})
and then the difference $\lambda_{H\phi}-\frac{\lambda_{HS}}{\lambda_{\phi S}}\lambda_\phi$ 
is not affected by it.} 
active neutrino masses can be generated through this process also in the present model. 
As is well known, they can be consistent with the neutrino oscillation data 
for such a small $\tilde\lambda_5$ even if $M_j$ and $M_\eta$ take TeV scale values.

For simplicity,  we assume that neutrino Yukawa couplings $h_{\alpha j}$ realize 
the tribimaximal mixing \cite{otribi} in the neutrino sector,\footnote{Although it cannot cause 
a nonzero mixing angle $\theta_{13}$ in the neutrino sector, a favorable value of $\theta_{13}$ 
 could be obtained in the PMNS matrix \cite{pmns} through the help of mixing in the charged 
lepton sector \cite{theta13,n1dmo}. }  
and we take them in the weak interaction basis as \cite{tribi0,tribi} 
\begin{equation}
h_{ej}=h_{\mu j}=-h_{\tau j}=\frac{1}{\sqrt 3}h_j e^{i\gamma_j}~(j=1,2);\qquad 
h_{e3}=0, ~h_{\mu 3}=h_{\tau 3}=\frac{1}{\sqrt 2}h_3e^{i\gamma_3}.
\label{ntribi}
\end{equation}
In that case, neutrino mass eigenvalues are found to be expressed as 
\begin{equation}
m_1= 0 ,  \qquad  m_2=  h_1^2\Lambda_1e^{2i\gamma_1}+h_2^2\Lambda_2e^{2i\gamma_2} ,   
\qquad   m_3=h_3^2\Lambda_3e^{2i\gamma_3}  
\end{equation}
This formula suggests that $N_1$ could be irrelevant to the 
neutrino mass eigenvalues for a sufficiently small $h_1$.
Data for the neutrino mass and mixing obtained from experiments 
and cosmological observations  suggest that $m_1$ is sufficiently small and 
the tribimaximal mixing is rather good.  
If the normal ordering, the hierarchical mass and the tribimaximal mixing are satisfied in
the neutrino sector,  eq.~(\ref{ntribi}) can be derived automatically and then
the qualitative conclusions obtained on the basis of this assumption could be robust
although details might be dependent on the numerical assumption.
However, we should note that the conclusions given here could be changed if these conditions are 
not satisfied in the neutrino sector.  

If we apply the RGE results obtained for the benchmark parameters given in eq.~(\ref{bench}) 
to this formula, the neutrino oscillation data \cite{pdg} are found to be explained by taking 
the neutrino Yukawa couplings $h_2$ and $h_3$ as
\begin{equation}
h_2= 4.24 \times 10^{-4}  , \qquad   h_3= 1.58\times 10^{-3}.
\label{h23}     
\end{equation} 
The RGE evolution of $y_{N_{2,3}}$ caused by $h_{2,3}$ is found to resolve the mass degeneracy 
of $N_2$ and $N_3$ at low energy regions as addressed in the previous part.

$N_1$ can be irrelevant to the result given in eq.~(\ref{h23}) when $h_1$ is small enough. 
We should also note that $h_1$ can take a larger value than $h_2$ without affecting it 
if $M_1\ll M_\eta$ is satisfied.  In that case, $\Lambda_1$ can be approximated as 
$\Lambda_1\simeq \frac{\tilde\lambda_5 v_H^2}{32\pi^2}\frac{M_1}{M_\eta^2}$, and then
it takes a heavily suppressed value compared with $\Lambda_2$. 
As a result, if the condition 
\begin{equation}
 h_1^2\left(\frac{M_1}{1~{\rm GeV}}\right)< 10^{-6}
\label{h1cond}
\end{equation} 
is satisfied, $h_2$ in eq.~(\ref{h23}) is not changed.
It suggests that $h_1>h_2$ is possible for $M_1<O(1)$ GeV in the present case.

It may also be useful to give a remark on the relation of this condition to a constraint imposed by
the lepton flavor violating process (LFV) $\mu\rightarrow e\gamma$, which gives an upper bound 
for $h_1$ \cite{lfv}. Its branching ratio can be expressed in the present model as \cite{mueg}
\begin{eqnarray}
&&Br(\mu\rightarrow e\gamma)=\frac{3\alpha}{64\pi(G_F M_{\eta^\pm})^2}\left|\sum_{j=1,2}
\frac{h_j^2e^{2i\gamma_j}}{3}F_2\left(\frac{M_j^2}{M_{\eta^\pm}^2}\right)\right|^2, \nonumber \\
&&F_2(x)=\frac{1-6x+3x^2+2x^3-6x^2\ln x}{6(1-x)^4}.
\end{eqnarray}
If we apply it to the model with $M_1\ll M_\eta$, the present experimental bound gives 
a constraint $h_1<0.12$ independently of $M_1$ since $F_2(x)$ converges to $\frac{1}{6}$
at $x\rightarrow 0$. Thus, this constraint should be taken care in the case $M_1~{^<_\sim}~100$~keV
since it could be stronger than the condition (\ref{h1cond}).   
In the next part, we discuss dark matter and leptogenesis taking account of these points.

\subsection{Dark matter}
The existence of a stable particle is guaranteed by the exact $Z_2$ symmetry. 
A possible DM candidate is either the lightest neutral component of $\eta$ or 
the lightest right-handed neutrino $N_1$ which has its odd parity. 
Since $M_{\eta_R}$ is smaller than $M_{\eta^\pm}$ and $M_{\eta_I}$ in the case 
$\tilde\lambda_4, \tilde\lambda_5~<0$ as found from eq.~(\ref{etamass2}),\footnote{
For the assumed parameters in  eq.~(\ref{bench}), we find that $\tilde\lambda_4<0$
 is satisfied at the relevant scale.}  $\eta_R$ is considered to be a DM candidate.
If reheating temperature is higher than the mass of the DM candidate and it could be 
in the thermal equilibrium, the DM abundance might be realized as their thermal relics 
through the freeze-out scenario \cite{freezeout}.
In fact, inflation is expected to be caused by the singlet scalar $S$ 
if the Planck mass is induced in some way 
and $S$ has a nonminimal coupling $\frac{\xi}{2} S^2R$ with the Ricci scalar $R$ \cite{sinfl}. 
It can be consistent with the CMB data of the Planck satellite \cite{planck}
if $\xi$ is fixed appropriately for the quartic coupling $\lambda_S$. 
Moreover, preheating following it can be expected to realize reheating temperature 
higher than $O(10^4)$ GeV \cite{sinfl,reheat}. It is high enough to make $\eta$ and $N_{2,3}$ 
in the thermal bath.  

A DM candidate $\eta_R$ is considered to be in the thermal equilibrium by the SM interactions.
On the other hand, $N_{2,3}$ could be in the thermal equilibrium through the 2-2 scatterings 
$\eta \eta^\dagger \rightarrow N_iN_j^c$ and $\ell_\alpha\ell_\beta^\dagger\rightarrow N_iN_j^c$
caused by the Yukawa couplings $h_2$ and $h_3$ given in eq.~(\ref{h23}), 
which explain the neutrino oscillation data.
The production of $N_1$ is expected to be proceeded through the scatteing 
$N_jN_j\rightarrow N_1N_1~(j=2,3)$ caused by $y_{N_j}$ \cite{nns} and also both the above 
scatterings and the decay of $\eta$ caused by $h_1$. 
Here, we consider the Boltzmann equations for the number density of $\eta_R$ and $N_1$ 
based on these processes. 
We express the number density of particle species $i$ in the comoving volume as 
$Y_i=\frac{n_i}{s}$ where $n_i$ and $s$ are the number density of the particle $i$ 
and the entropy density of the Universe, respectively.
If we define a dimensionless parameter $z$ for a temperature $T$ as $z\equiv \frac{M_2}{T}$ 
and represent an equilibrium value of $Y_i$ as $Y_i^{\rm eq}$, 
the Boltzmann equations for $Y_{\eta_R}$ and $Y_{N_1}$ can be written as 
\begin{eqnarray}
\frac{dY_{\eta_R}}{dz}&=&-\frac{z}{H(M_2)s}\left[\sum_{a,b=\eta_R,\eta_I,\eta^\pm} \gamma^S(ab)
\left(\frac{Y_{\eta_R}^2}{Y_{\eta_R}^{{\rm eq}2}}-1\right)
+\gamma^D(\eta_R)\left(\frac{Y_{\eta_R}}{Y_{\eta_R}^{\rm eq}}-1\right)\right],   \label{bolt1}
\label{yeta} \\
\frac{dY_{N_1}}{dz}&=&-\frac{z}{H(M_2)s}\left[
\left\{\sum_{j=2,3}\gamma_1^S(N_jN_j)+\sum_{a=\eta,\ell}\gamma_1^S(aa^\dagger)\right\}
\left(\frac{Y_{N_1}^2}{Y_{N_1}^{{\rm eq}2}}-1\right)\right.  \nonumber \\
&-&\left.\gamma^D(\eta_R)\left(\frac{Y_{\eta_R}}{Y_{\eta_R}^{\rm eq}}-1\right)\right],
\label{bolt2}
\end{eqnarray}
where $H(M_2)$ is the Hubble parameter at $T=M_2$.
Reaction density of the decay $\eta_R\rightarrow \nu N_2$ is represented by 
$\gamma^D(\eta_R)$, and reaction densities of the scatterings 
$\ell\ell^\dagger\leftrightarrow N_j^cN_j$, $\eta\eta^\dagger\leftrightarrow N_j^cN_j$ and $N_iN_i\leftrightarrow N_jN_j$ are represented by
$\gamma^S_j({\ell\ell^\dagger})$, $\gamma^S_j(\eta\eta^\dagger)$ and $\gamma_j^S(N_iN_i)$. 
They are respectively related to the thermally averaged decay width 
$\langle\Gamma_a\rangle$ for $a\rightarrow ij$ and the thermally averaged cross 
section $\langle\sigma_{ab}|v|\rangle$ for the scattering $ab\leftrightarrow ij$ as\cite{luty} 
\begin{equation}
\gamma^D(a)=\langle\Gamma_a\rangle n_a^{\rm eq}, \qquad
\gamma^S(ab)=\langle\sigma_{ab}|v|\rangle n_a^{\rm eq}n_b^{\rm eq},
\end{equation} 
where $n_a^{\rm eq}$ is the equilibrium number density of a particle $a$.

In the $\eta_R$ mass region considered in this study, which is so called high mass region 
($M_{\eta_R}~{^>_\sim}~500$ GeV) in
the inert doublet model, the couplings $\tilde\lambda_3$ and $\tilde\lambda_4$ play a crucial role 
to reduce its relic abundance through the coannihilation among the components of $\eta$ \cite{etadm}.
It is taken into account in $\gamma^S(ab)$ of eq.~(\ref{yeta}).
Unfortunately, since the absolute values of $\tilde\lambda_3$ and $\tilde\lambda_4$ 
in this model are much smaller than the required ones, 
the relic $\eta_R$ is difficult to be reduced to the observed DM abundance.\footnote{Moreover, 
we find that the couplings in the scalar potential $V_0$ reach the Landau pole at a scale 
like $O(10^8)$ GeV much lower than the Planck scale if $\tilde\lambda_3$ and $\tilde\lambda_4$ 
have magnitude which can reduce the $\eta_R$ abundance sufficiently. 
It means that the present framework cannot be consistent with the $\eta_R$ DM in this mass region.}
As a result, we have to consider the lightest right-handed neutrino $N_1$ as DM. In that case, 
$M_1<M_{\eta_R}$ has to be satisfied. It does not bring any contradiction between the present 
scheme for the neutrino mass generation and the neutrino oscillation data as found 
from the discussion on the neutrino mass. 
The lightest right-handed neutrino has been studied as a DM candidate from various view points \cite{lfv,tribi0,tribi,n1dmv,n1dms,n1dmo}.

We define $Y_{N_1}^\infty$ as the $N_1$ relic density in the present Universe which satisfies 
$\Omega_{N_1}h^2=0.12$. We also define $Y_{\eta_R}^f$ as a value of $Y_{\eta_R}$ at a temperature 
where the $\eta_R$ coannihilation by the electroweak gauge interaction freezes out.
The fact that $\eta_R$ finally decays to $N_1$ requires that they satisfy 
$Y_{\eta_R}^f\le Y_{N_1}^\infty$. 
Since the solution of the above Boltzmann equations shows that $Y_{\eta_R}^f\simeq Y_{N_1}^\infty$ is 
realized for $M_1\simeq 45$ GeV, we find that  $M_1~{^<_\sim}~45$ GeV should be satisfied as an
upper bound of $M_1$. Within this range of $M_1$, $N_1$ could be 
produced to reach the thermal equilibrium through the processes caused by $h_1$ and $y_{N_j}$.
In that case, however, its annihilation described by $\gamma_1^S(\ell\ell^\dagger)$ and 
$\gamma_1^S(\eta\eta^\dagger)$ is suppressed due to $M_1\ll M_\eta$ and then the 
$N_1$ abundance is difficult to be reduced to the required level for  the bound of $h_1$ discussed 
in the last part of the previous section.\footnote{Coannihilation among the right-handed 
neutrinos might solve this difficulty \cite{tribi}. 
However,  $N_1$ does not have a degenerate mass with $N_{2,3}$ in the present model.}    
Although the freeze-out scenario is difficult to be applied to $N_1$ in this model,  
the freeze-in scenario \cite{freezein} could work well for $N_1$ to realize the 
suitable relic abundance. 

Freeze-in of $N_1$ is caused through the processes addressed above, which are caused by the 
couplings $y_{N_1}$ and $h_1$. 
Here, we should note that $M_1(\equiv y_{N_1}v_\phi)$ determines both 
the $N_1$ production due to the scattering $N_jN_j\rightarrow N_1N_1$ represented by
$\gamma^S_1(N_jN_j)$ in eq.~(\ref{bolt2}) and a value of $Y_{N_1}^\infty$.
The former is proportional to $y_{N_1}^2$ and the latter is proportional to $y_{N_1}^{-1}$.
If this scattering process for a certain $M_1$ value results in $Y_{N_1}>Y_{N_1}^\infty$, 
the $N_1$ abundance cannot be reduced to $Y_{N_1}^\infty$ since the $N_1$ annihilation is ineffective 
just as in the freeze-out case.
This means that there is an upper bound for $M_1$ which is fixed by $Y_{N_1}=Y_{N_1}^\infty$.
It is much smaller than the one obtained from the condition $Y_{\eta_R}^f\sim Y_{N_1}^\infty$. 
On the other hand, although this scattering process for a smaller $M_1$ results in 
$Y_{N_1}<Y_{N_1}^\infty$, the decay of the $\eta_R$ in the thermal bath could 
help $Y_{N_1}$ to reach $Y_{N_1}^\infty$ if $h_1$ takes a suitable value.
For much smaller $M_1$, the $N_jN_j$ scattering becomes ineffective to generate $N_1$. 
However, in that case, 
both the scattering processes and the $\eta_R$ decay caused by $h_1$ could make 
$Y_{N_1}$ reach $Y_{N_1}^\infty$ since $h_1$ can take a larger value within the 
constraints derived from eq.~(\ref{h1cond}) and the LFV.   
These are considered to be general features found for the $N_1$ DM 
in this model independently of details of parameters assumed here. 

\begin{figure}[t]
\vspace*{-1cm}
\begin{center}
\includegraphics[width=10cm]{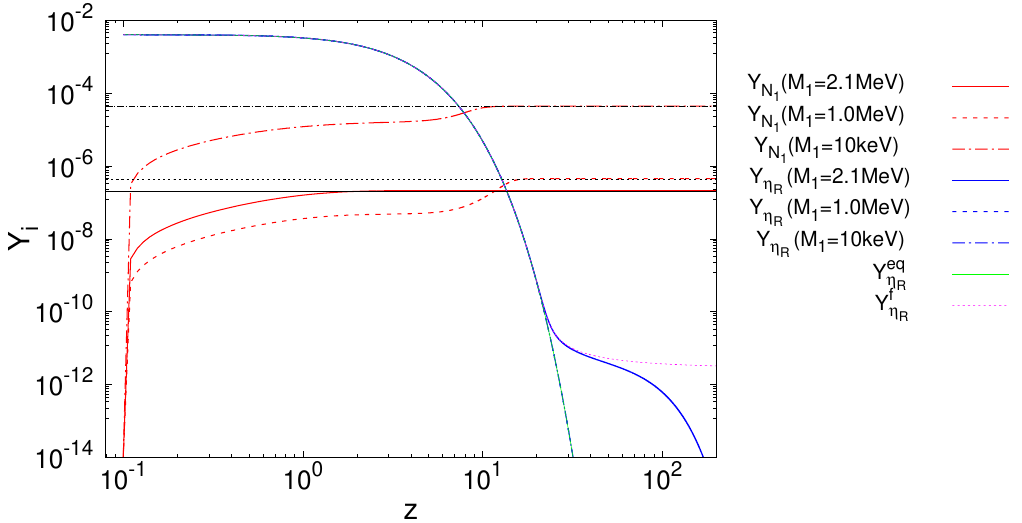}
\end{center}
\vspace*{-5mm}
\footnotesize{{\bf Fig.~3}~~~Evolution of the abundance of $N_1$ and $\eta_R$ in the comoving 
volume is drawn as a function of $z(\equiv M_2/T)$. As initial values at $z=0.1$, 
$Y_{\eta_R}=Y_{\eta_R}^{\rm eq}$ and $Y_{N_1}=0$ are assumed. 
$Y_{N_1}$ is plotted by red lines for three cases of $(M_1, h_1)$; 
$( 2.1~{\rm MeV}, 10^{-8})$ (solid), $( 1.0~{\rm MeV}, 1.3\times 10^{-5})$ (dashed) and 
$(10~{\rm keV}, 2.8\times 10^{-4})$ (dash-dotted).
Horizontal black lines represent $Y_{N_1}^\infty$ which is required to realize 
$\Omega_{N_1}h^2=0.12$ for each mass. The same type of lines are used as the ones 
of $Y_{N_1}$ for the corresponding mass. $Y_{\eta_R}$ is plotted by blue lines for each $M_1$ value. 
In latter two cases, $Y_{\eta_R}=Y_{\eta_R}^{\rm eq}$ is realized through the processes caused by
$h_1$. Only in the $M_1=2.1$ MeV case, $Y_{\eta_R}$ is found to leave the equilibrium 
due to a small $h_1$ value.}
\end{figure}

In Fig.~3, we plot $Y_{N_1}$ obtained as a solution of the Boltzmann equations for three 
typical cases defined by $(M_1, h_1)$. They correspond to three possibilities described above, respectively. 
In this calculation, $M_{2,3}$ and $M_{\eta_R}$ are fixed using the values given in eq.~(\ref{mscalar}). 
The upper bound of $M_1$ is found to be 2.1 MeV for which the $N_jN_j$ scattering 
realizes $Y_{N_1}^\infty$ by itself. 
In the $M_1=1.0$ MeV case, the $N_jN_j$ scattering produces $N_1$ but its abundance 
cannot reach $Y_{N_1}^\infty$. However,  the $N_1$ produced through the decay of $\eta_R$ 
in the thermal bath supplies the deficit abundance of $N_1$ to make it reach $Y_{N_1}^\infty$ 
if $h_1$ is fixed to a suitable value within the allowed range. 
For much smaller values of $M_1$, the processes caused by a larger value of $h_1$ 
becomes dominant ones. We can find it from the $Y_{N_1}$ evolution in the case $10$ keV. 
From these results, we can conclude that
the model predicts a right-handed neutrino DM in the mass range less than $O(1)$ MeV.  
It is interesting that this range includes the mass of the sterile neutrino for which
it could give an explanation for the small scale structure problem \cite{wdm}.
It should be noted that this sterile neutrino has no mixing with active neutrinos. 
Although numerical values like the upper bound of $M_1$ 
depend on the values of parameters given in eq.(\ref{bench}), 
the features found here are expected to be rather general for this framework 
and the right-handed neutrino could be light DM with mass less than a GeV scale. 

Finally, we give a brief comment on a possibility such that the lightest 
neutral component of $\eta$ is DM and $M_\eta<M_1$ is satisfied in this framework 
which is featured by the small couplings $\tilde\lambda_{3,4,5}$. 
In the present work, we confine our study to the case where $\eta_R$ mass is in the high mass region.
However, if its mass is in the low mass region ($^<_\sim ~80$ GeV) or in the intermediate mass region 
($80 - 200$ GeV), its relic abundance is known to be realized through the SM interaction \cite{etadm}
although their allowed parameter space is severely constrained through various 
experiments like the direct detection \cite{direct} by now.    
Since $M_\eta<M_j$ breaks the assumption adopted here,
we need to study whether the weak scale can be generated successfully in such cases.
It is an interesting subject in this model but it is beyond the scope of the present study. 

\subsection{Baryon number asymmetry}
In this model, the mass ordering $M_1<m_\eta< M_2\simeq M_3$ is satisfied for the $Z_2$ odd 
contents as shown in eq. (\ref{mscalar}) and the decay width $\Gamma_j$ of $N_j$ is found to be 
$\Gamma_2<\Gamma_3$.
It suggests that baryon number asymmetry is expected to be produced through leptogenesis 
caused by the out-of-equilibrium decay of $N_2$ 
to $\eta$.\footnote{We confine our study to the thermal leptogenesis and do not consider 
a possibility of nonthermal leptogenesis \cite{ext}.} 
Almost degenerate masses of $N_2$ and $N_3$ realized in the model are expected to 
enhance the $CP$ asymmetry $\varepsilon$ in the $N_2$ decay so that a value of $\varepsilon$ 
could be large enough for 
the generation of the required lepton number asymmetry. 
It can be expressed by using eq.~(\ref{ntribi}) as \cite{resonant}
\begin{eqnarray}
\varepsilon&=&\frac{ {\rm Im}
\left[\left(\sum_\alpha h_{\alpha 2}h^\ast_{\alpha 3}\right)^2\right]}
{ (\sum_\alpha h_{\alpha 2}h_{\alpha 2}^\ast)
(\sum_\alpha h_{\alpha 3}h_{\alpha 3}^\ast)}
\frac{(M_2^2-M_3^2)M_2\Gamma_{N_3}}{(M_2^2-M_3^2)^2+M_2^2\Gamma_{N_3}^2} \nonumber \\
&=&\frac{2}{3}\frac{(M_2^2-M_3^2)M_2\Gamma_{N_3}}{(M_2^2-M_3^2)^2+M_2^2\Gamma_{N_3}^2},
\label{res}
\end{eqnarray} 
where we assume $\sin 2(\gamma_2-\gamma_3)=1$ in the second line.
If we define $\delta\equiv (M_2-M_3)/M_2$ as a parameter to represent the mass degeneracy
between $N_2$ and $N_3$, the parameters given 
in eq.~(\ref{bench}) realizes $\delta=O(10^{-6})$ at the scale $M_2$ and then
$\varepsilon$ takes a value of $O(10^{-5})$.   

In the study of leptogenesis in the present model with $M_2\simeq M_\eta$,
we should note that the lepton number asymmetry and the $N_1$ abundance could have a 
close relation since $\eta$ produced in the $N_2$ decay 
can contribute to the $N_1$ abundance through the $\eta$ decay.
If we take account of this point, we have to solve the Boltzmann equations for 
$\eta_R, N_1,N_2, N_3$ and the lepton number asymmetry $Y_L$ as
coupled equations. As found from Fig~3, the freezeout of the coannihilation of $\eta$
occurs around $z\simeq 25$, and then the $\eta$ produced through the $N_2$ decay 
at $z~{^<_\sim}~25$ is expected to be in the thermal equiliburium. This situation is the same 
in the present case except that the additional $\eta$ is supplied in the thermal bath.
Thus, when we take account of the effect of $N_2$ decay, we have to assume a smaller 
value for $h_1$ in order to realize the required $N_1$ relic abundance compared with the one used 
in Fig.~3. In this point, $M_1=2.1$ MeV is an exceptional case since $N_1$ is generated 
only through the scatterings caused by $y_{N_j}$.
We can confirm it by solving the Boltzmann equations.  
As a result, when we consider leptogenesis in this model, 
the equations for $Y_{N_{2,3}}$ and $Y_L$ can be treated separately from the ones 
for $Y_{\eta_R}$ and $Y_{N_1}$.
Three cases given in Fig.~3 can be treated as the same case in the study of leptogenesis 
since the relevant equations are independent of $M_1$ and $h_1$.

The Boltzmann equations for them are given as 
\small
\begin{eqnarray}
&&\frac{dY_L}{dz}=\frac{-z}{H(M_2)s}\left[-\varepsilon \gamma^D(N_2)
\left(\frac{Y_{N_2}}{Y_{N_2}^{\rm eq}}-1\right)
+\left\{\sum_{k=2,3}\gamma^D(N_k)+ 2\Big(\gamma^S(\ell \eta^\dagger)
+\gamma^S(\ell\ell)\Big) \right\}\frac{Y_L}{Y_\ell^{\rm eq}}\right], \nonumber \\
&&\frac{dY_{N_3}}{dz}=\frac{- z}{H(M_2)s}\left[\left\{\sum_{a=\eta,\ell}\gamma^S_3(aa^\dagger)  
+\gamma_3^S(N_2N_2)\right\}\left(\frac{Y_{N_3}^2}{Y_{N_3}^{{\rm eq}2}}-1\right) +\gamma^D(N_3)\left(\frac{Y_{N_3}}{Y_{N_3}^{\rm eq}}-1\right)\right], 
\nonumber \\
&&\frac{dY_{N_2}}{dz}=\frac{- z}{H(M_2)s}\left[\left\{\sum_{a=\eta,\ell}\gamma^S_2(aa^\dagger) 
+\gamma_2^S(N_3N_3)\right\}\left(\frac{Y_{N_2}^2}{Y_{N_2}^{{\rm eq}2}}-1\right) +\gamma^D(N_2)\left(\frac{Y_{N_2}}{Y_{N_2}^{\rm eq}}-1\right)\right], 
\nonumber \\
\label{bolt}
\end{eqnarray}
\normalsize
where we take account of $Y_{N_2}^{\rm eq}\simeq Y_{N_3}^{\rm eq}$.
$\gamma^S(\ell\eta^\dagger)$ and $\gamma^S(\ell\ell)$ 
are reaction densities of the lepton number violating scatterings 
$\ell\eta^\dagger \leftrightarrow \ell^\dagger\eta$ and $\ell\ell\leftrightarrow \eta\eta$.
We solve them for an initial condition such that $Y_{N_j}=Y_L=0$ at
the reheating temperature $T_R$. 

\begin{figure}[t]
\vspace*{-1cm}
\begin{center}
\includegraphics[width=10cm]{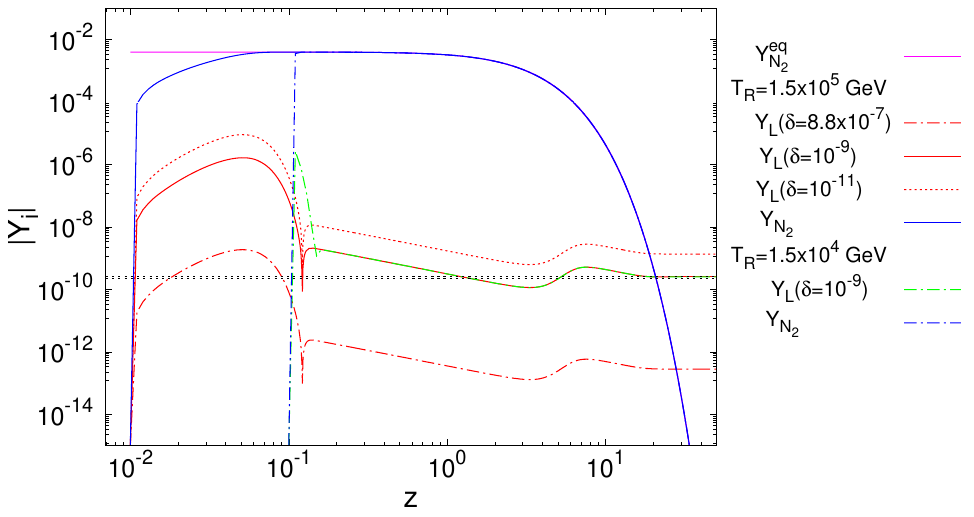}
\end{center}
\vspace*{-5mm}
\footnotesize{{\bf Fig.~4}~~~Evolution of  $Y_{N_2}$ and $Y_L$ generated by the $N_2$ 
decay. $Y_L$ is plotted for three values of $\delta$ assuming
$T_R=1.5\times 10^5$ GeV. Horizontal dashed lines represent a region of $Y_L$ required 
for the explanation of the baryon number asymmetry in the Universe. 
In the case  $T_R=1.5\times 10^4$ GeV,  $Y_L$ is also plotted for the case 
$\delta=10^{-9}$, which coincides with the one obtained in the case $T_R=1.5\times 10^5$ GeV. 
The behavior of $Y_L$ around $z\simeq 10$ is considered to be caused by the $N_2$ 
generation due to $\gamma_2^S(N_3N_3)$ and the $Y_L$ washout due to 
$\gamma^S(\ell\ell)$ whose magnitudes are reversed there.  }
\end{figure}

The solutions of these Boltzmann equations are shown in Fig.~4.\footnote{The baryon number 
asymmetry is estimated by using the lepton number asymmetry at $z \simeq 15$ since 
the sphaleron is in the thermal equilibrium at $z~{^<_\sim}~15$ for the $M_2$ assumed here. } 
In this calculation,  we apply the same parameters used in Fig.~3.
We find that the lepton number asymmetry $Y_L$ sufficient for the explanation of 
the baryon number asymmetry in the Universe cannot be generated for th mass degeneracy 
$\delta$ derived by using $y_{N_{2,3}}$ assumed in eq.~(\ref{bench}). 
This $\delta$ is not small enough to make the $CP$ asymmetry $\varepsilon$ large enough
since the washout processes are not sufficiently suppressed around $z\simeq 1$.
To improve this result, we may consider the realization of stricter mass degeneracy 
between $N_2$ and $N_3$.
We modify the relation between $y_{N_2}$ and $y_{N_3}$ given in eq.~(\ref{bench}) to
\begin{equation}
y_{N_3}=(1+\delta_0)y_{N_2}, \qquad y_{N_2}=0.203.
\end{equation}
If $\delta_0$ is tuned appropriately, $\delta$ at the scale $M_2$ can be smaller than $O(10^{-9})$. 
In that case, $\varepsilon$ can take a value larger than $O(10^{-3})$ while other results are kept  
unchanged.
In fact, the sufficient lepton number asymmetry can be generated as found in Fig.~4.
Although initial values of parameters have to be tuned at the Planck scale, 
promising low energy results can be obtained in a consistent way. 
It is noticeable that the mass degeneracy among $N_{2,3}$ and $\eta_R$ which is imposed for the  
realization of the appropriate electroweak symmetry breaking 
could play an important role also for the successful leptogenesis.
 
\section{Summary}
The scotogenic model is a promising simple extension of the SM at TeV scales.  
It can explain the neutrino mass, the DM existence, and the baryon number asymmetry in the
Universe which are unsolvable problems in the SM.
This model has three mass scales such that the electroweak scale, 
the masses of the inert doublet scalar and the right-handed neutrinos. 
The latter two are usually assumed to be in TeV regions.
However, the model cannot explain both the origin of these mass scales and their stability against 
the quantum correction just as the SM.  

In this paper, in order to solve this problem, the model is extended by imposing the classical 
scale invariance and the custodial symmetry $SO(5)$. 
By introducing the two real scalars, the scalar potential is modified to be invariant 
under these symmetries. An intermediate scale is induced through the spontaneous breaking of 
these symmetries caused by the Colman-Weinberg mechanism.
Four Nambu-Goldstone bosons are brought about as a result of the custodial symmetry breaking 
$SO(5)\rightarrow SO(4)$. Three of them are eaten by the weak gauge boson and 
a remaining one becomes the Higgs-like scalar.  It has small mixing with a dilaton which appears 
along with the violation of the scale invariance. 
Since the negative squared Higgs mass generated from the intermediate scale is suppressed by the
custodial symmetry, the weak scale can be much smaller than the intermediate scale.

The imposed symmetries constrain the scalar potential of the ordinary scotogenic model 
which is included as a part of this extended model. 
This constrained scalar potential and the existence of the singlet scalars 
could cause phenomenological consequences different from the ones of the ordinary scotogenic
model. In this study,  a certain texture of the neutrino Yukawa couplings is assumed for the analysis. 
Using such an assumption, we find that a DM candidate has to be not the lightest neutral component of 
the inert doublet scalar $\eta$ but the lightest right-handed neutrino. 
Moreover, the freezeout scenario cannot be applied for it to explain the DM abundance and then
only the freezein scenario is applicable for it. 
Since the right-handed neutrino mass is induced through the coupling to a singlet scalar,
the lightest right-handed neutrino as DM is produced through the scattering among 
right-handed neutrinos mediated by it. This nature severely constrains its mass in the range
less than $O(1)$ MeV. 
It contains a mass region where the small scale structure problem could be resolved partially
through a sterile neutrino.
The baryon number asymmetry is found to be explained through the resonant leptogenesis. 
An interesting point of the model is that the mass degeneracy among $N_{2,3}$ and $\eta$
assumed for the appropriate generation of the electroweak scale also plays a crucial 
role to make the model consistent with the successful leptogenesis through the resonance. 
Although further study is required to clarify the phenomenological consequences 
in the cases defined by more general neutrino Yukawa textures, the model with a suitable neutrino 
Yukawa texture may be considered as a candidate of the model which can solve the problems 
in the scotogenic model.      
 
\section*{Appendix}
We use one-loop RGEs in this study for simplicity.
RGEs for the scalar quartic coupling constants in eq.~(\ref{spot}) are given as
\small
\begin{eqnarray}
\beta_{\lambda_H}&=&\frac{1}{16\pi^2}\left[24\lambda_H^2+2\lambda_{H\phi}^2
+\frac{1}{2}\lambda_{HS}^2
-6y_t^4+\frac{3}{8}\left(g_Y^2+g_2^2\right)^2+\frac{3}{4}g_2^4
+\lambda_H\left(12y_t^2-3g_Y^2-9g_2^2\right) \right], \nonumber \\
\beta_{\lambda_{H\phi}}&=&\frac{1}{16\pi^2}\left[6\lambda_{H\phi}\lambda_\phi
+12\lambda_H\lambda_{H\phi} +8\lambda_{H\phi}^2+\frac{1}{2}\lambda_{\phi S}\lambda_{HS}
+\frac{1}{2} \lambda_{H\phi}\left(12y_t^2+4\sum_jy_{N_j}^2-3g_Y^2-9g_2^2\right)\right], 
\nonumber \\
\beta_{\lambda_\phi}&=&\frac{1}{16\pi^2}\left(18\lambda_\phi^2+8\lambda_{H\phi}^2
+\frac{1}{2}\lambda_{\phi S}^2 +2\lambda_{\eta\phi}^2
-4\sum_jy_{N_j}^4+4\lambda_\phi \sum_jy_{N_j}^2\right), \nonumber \\ 
\beta_{\lambda_{HS}}&=&\frac{1}{16\pi^2}\left[12\lambda_H\lambda_{HS}+
2\lambda_{H\phi}\lambda_{\phi S}+2\lambda_S\lambda_{HS}+4\lambda_{HS}^2
+\frac{1}{2}\lambda_{HS}\left(12h_t^2-3g_Y^2-9g_2^2\right)\right], \nonumber \\
\beta_{\lambda_{\phi S}}&=&\frac{1}{16\pi^2}\left(6\lambda_{\phi S}\lambda_\phi
+8\lambda_{H\phi}\lambda_{HS}+2\lambda_S\lambda_{\phi S}+4\lambda_{\phi S}^2
+2\lambda_{\phi S}\sum_j y_{N_j}^2\right),  \nonumber \\
\beta_{\lambda_S}&=&\frac{1}{16\pi^2}\left(3\lambda_S^2+3\lambda_{\phi S}^2
+12 \lambda_{HS}^2 \right) \nonumber \\
\beta_{\lambda_\eta}&=&\frac{1}{16\pi^2}\left[24\lambda_\eta^2+\frac{1}{2}\lambda_{\eta\phi}^2
+\frac{3}{8}\left(g_Y^2+g_2^2\right)^2+\frac{3}{4}g_2^4-\lambda_\eta\left(3g_Y^2+9g_2^2\right)
\right], \nonumber \\
\beta_{\lambda_{\eta\phi}}&=&\frac{1}{16\pi^2}\lambda_{\eta\phi}\left[6\lambda_\phi
+12\lambda_\eta +4\lambda_{\eta\phi}
+\frac{1}{2}\left(4\sum_jy_{N_j}^2-3g_Y^2-9g_2^2\right)\right],  \nonumber \\
\beta_{\tilde\lambda_5}&=&\frac{1}{16\pi^2}\tilde\lambda_5
\left[4\lambda_H+4\lambda_\eta+\left(6h_t^2-3g_Y^2-9g_2^2\right)\right].
\label{eq1}
\end{eqnarray}
\normalsize
In these RGEs, we take account of the top Yukawa coupling $y_t$, the neutrino Yukawa 
couplings $h_j$ and $y_{N_j}$ among the Yukawa couplings in the model.
One-loop RGEs for $h_j$ and $y_{N_j}$ are given as
\begin{eqnarray}
\beta_{h_j}&=&\frac{h_i}{16\pi^2}\left[\frac{5}{2}(h_1^2+h_2^2)+h_3^2+\frac{1}{2}y_{N_i}^2-\frac{9}{4}g_2^2
+\frac{1}{4}g_Y^2\right] \quad (i=1,2),  \nonumber \\
\beta_{h_3}&=&\frac{h_3}{16\pi^2}\left[h_1^2+h_2^2+\frac{5}{2}h_3^2+\frac{1}{2}y_{N_3}^2-\frac{9}{4}g_2^2
+\frac{1}{4}g_Y^2\right], 
 \nonumber \\
\beta_{y_{N_j}}&=&\frac{y_{N_j}}{16\pi^2}\left(3y_{N_j}^2+2\sum_{k=1}^3y_{N_k}^2+ 2h_j^2\right) \quad (j=1,2,3).
\label{eq3}
\end{eqnarray}
It should be noted that the $\beta$-functions of the scalar quartic couplings indirectly depend on 
the neutrino Yukawa couplings $h_j$ through $y_{N_j}$. Thus, when $N_1$ is very light and then 
$y_{N_1}$ is sufficiently small, the evolution of these quartic couplings is almost independent of $h_1$ 
as long as the condition (\ref{h1cond}) is satisfied.  

Scalar quartic couplings 
\begin{equation}
\frac{1}{2}\lambda_{\eta S}S^2(\eta^\dagger\eta)+
\tilde\lambda_3(H^\dagger H)(\eta^\dagger\eta)+\tilde\lambda_4(H^\dagger\eta)(\eta^\dagger H)
\end{equation}
can be induced radiatively through their RGEs, which are given as
\begin{equation}
\beta_{\lambda_{\eta S}}=\frac{1}{16\pi^2}\lambda_{\phi S}\lambda_{\eta\phi},  \qquad
\beta_{\tilde\lambda_3}=\frac{1}{16\pi^2}\left(\lambda_{H\phi}\lambda_{\eta\phi}+\tilde\lambda_5^2
\right),  \qquad
\beta_{\tilde\lambda_4}=\frac{1}{16\pi^2}\tilde\lambda_5^2.
\label{eq2}
\end{equation}
In the $\beta$-functions given in eqs.~(\ref{eq1}) and (\ref{eq2}), terms proportional 
to  $\lambda_{\eta S}$, $\tilde\lambda_3$, and $\tilde\lambda_4$ are not explicitly written
since they are assumed to be zero at the Planck scale. 
Although they give much smaller contributions than other couplings, 
they are taken into account in the present numerical calculation. 
We confirm that absolute values of $\tilde\lambda_3$ and $\tilde\lambda_4$ are of 
$O(10^{-6})$ at TeV regions for the benchmark parameters.
It justifies our treatment of these couplings in the present study.

\newpage

\end{document}